\documentclass[aps,prl,final,twocolumn,showpacs,superscriptaddress]{revtex4}

\usepackage{graphicx}
\usepackage{amsmath}
\usepackage{amsfonts}
\usepackage{natbib}

\newcommand{\drain}{\ensuremath{^1\!S_0\,-\,^3\!P_1}}

\newcommand{\mean}[1]{\ensuremath{\langle#1\rangle}}

\newcommand{\SYRTE}{\affiliation{LNE-SYRTE, Observatoire de Paris, CNRS, UPMC~; 61
avenue de l'Observatoire, 75014 Paris, France}}

\begin{document}

\title{Lattice Induced Frequency Shifts in Sr Optical Lattice Clocks at the $10^{-17}$ Level}

\author{P. G. Westergaard} \SYRTE
\author{J. Lodewyck} \SYRTE
\author{L. Lorini} \SYRTE
\affiliation{Istituto Nazionale di Ricerca Metrologica (INRIM), Strada delle Cacce
91, 10135 Torino, Italy}
\author{A. Lecallier} \SYRTE
\author{E. A. Burt} \SYRTE
\affiliation{Jet Propulsion Laboratory, California Institute of Technology
Pasadena, CA 91109-8099, U.S.A.}
\author{M. Zawada} \SYRTE
\affiliation{Institute of Physics, Nicolaus Copernicus University, Grudziadzka 5,
PL-87-100 Torun, Poland}
\author{J. Millo} \SYRTE
\author{P. Lemonde} \SYRTE

\begin{abstract}
We present a comprehensive study of the frequency shifts associated with the lattice potential for a Sr lattice clock. By comparing two such clocks with a frequency stability reaching $5\times 10^{-17}$ after a one hour integration time, and varying the lattice depth up to $U_0=900 \, E_r$ with $E_r$ being the recoil energy, we evaluate lattice related shifts with an unprecedented accuracy. We put the first experimental upper bound on the recently predicted frequency shift due to the magnetic dipole (M1) and electric quadrupole (E2) interactions. This upper bound is significantly smaller than the theoretical upper limit. We also give a new upper limit on the effect of hyperpolarizability with an improvement by more than one order of magnitude. Finally, we report the first observation of the vector and tensor shifts in a lattice clock. Combining these measurements, we show that all known lattice related perturbation will not affect the clock accuracy down to the $10^{-17}$ level, even for very deep lattices, up to $U_0=150\,E_r$.
\end{abstract}

\pacs{06.30.Ft, 42.62.Fi, 37.10.Jk, 42.50.Nn}
\maketitle

Together with single trapped ion clocks, optical lattice clocks represent the future of frequency metrology. These two types of apparatus now outperform microwave standards thanks to their five orders of magnitude higher clock transition frequency and to a stringent confinement-based control of atomic motion. In trapped ion clocks, one takes advantage of the external charge to confine atoms in the relatively weak electromagnetic RF fields of a Paul trap~\cite{Leibfried03}. In contrast, optical lattice clocks~\cite{Takamoto05} use neutral atoms and the trap based perturbations are orders of magnitude larger. Ion clocks currently exhibit the best accuracy at a level close to $10^{-17}$ in fractional units~\cite{Rosenband08,Chou10}. The best reported lattice clock accuracy is presently slightly above $10^{-16}$ with large room for improvement once one refines the control of the blackbody radiation shift and of lattice related effects~\cite{Ludlow08,Lemke09}. In terms of frequency stability, lattice clocks are expected to outperform their single ion counterpart thanks to the larger number of atoms available in the experiment.

The effects of the trapping potential are one of the main objects of study in lattice clocks. Indeed, at the theoretical minimal lattice depth $U_0 \sim 10\,E_r$ ($E_r$ being the recoil energy associated with the absorption of a lattice photon) required for effectively cancelling motional effects~\cite{Lemonde05}, the fractional energy shift of both clock transition levels is on the order of $10^{-10}$. In all practical experiments to date, $U_0$ is at least 5 to 10 times larger. Though the associated linear shift of the clock frequency is dramatically rejected by tuning the lattice to the magic wavelength\,\cite{KatoPal03}, residual higher order effects have been predicted as potential limitations to the clock accuracy~\cite{KatoPal03,Brusch06,Taichenachev08}. In addition, polarization dependent effects (vector and tensor shift, see below) have also been considered as possible serious issues for operating clocks with a 3D lattice with fermionic isotopes~\cite{Boyd072,akatsuka_optical_2008,swallows_suppression_2010}, while this configuration  is seen as the ultimate way to suppress the frequency shift due to collisions between cold atoms~\cite{akatsuka_optical_2008}. Finally, a better control of the lattice induced effects would allow operating the clock at larger $U_0$ than the theoretical minimum. This would make the clock operation easier, and more importantly, it would also allow for a better control of several systematic effects. For instance, the shift due to cold collisions between atoms is expected to critically depend on $U_0$ via confinement strength and site-to-site tunneling~\cite{Campbell09,Gibble09,gibble_frequency_2010,swallows_suppression_2010}.

In this letter, we address these problems with a detailed study of all known lattice related effects. We show that they can be controlled at the $10^{-17}$ level for a trap depth up to $U_0=150\,E_r$. Based on these measurements, we also propose an optimal 3D lattice configuration for the fermionic $^{87}$Sr isotope.


Three different lattice induced effects have been identified. The first one is related to the non-scalar feature of the atom-lattice interaction, which results from the atomic hyperfine structure~\cite{KatoPal03,Boyd072}. This structure induces the minute dipole electric moment of the clock transition allowing its laser excitation with extremely small linewidth. It also induces a small vector and tensor component in the atomic polarizability. In addition to the intensity dependence, this makes the light shift slightly dependent on the lattice polarization and geometry. The effect can be compensated for by tuning the lattice frequency, but this would be sensitive to possible long term variations of the laser's polarization and alignment as well as of the magnetic field direction. These effects have never been observed in a Sr lattice clock so far. The second effect, Hyperpolarizability, refers to a shift due to two-photon transitions. It was already pointed out in 2003~\cite{KatoPal03}. The frequency shift scales as $U_0^2$ and therefore cannot be compensated for by a change of the lattice frequency. A first experimental study showed that this effect was small enough to not alter Sr clock accuracy at the $10^{-18}$ level for $U_0=10\,E_r$~\cite{Brusch06}. A better knowledge of hyperpolarizability is still required to ensure full performance at larger $U_0$. Finally, a subtle effect of higher order multipolar terms has recently been predicted~\cite{Taichenachev08}. Dipole magnetic (M1) and quadrupole electric (E2) interactions indeed lead to a shift of the internal atomic energy levels that is linear in lattice laser intensity. However, their spatial dependence in a lattice formed by the interference of several traveling waves does not match the one of the main dipole electric term. This in turn leads to a change of the oscillation frequency in the lattice potential wells and thereby of the spacing of the atomic motional states. This effect is expected to differ for the two clock states. The net M1/E2 frequency shift of the clock frequency scales as $U_0^{1/2}$ and again cannot be compensated for by tuning the lattice frequency. In Ref.~\cite{Taichenachev08} it is argued that this effect could be as large as $10^{-16}$ for $U_0=50\,E_r$, a level at which the accuracy budget reported in~\cite{Ludlow08} would have to be amended.

\begin{figure}
\begin{center}
	\includegraphics[width=\columnwidth]{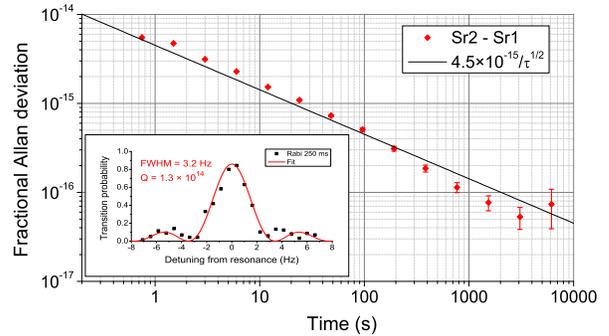}
	\caption{Allan deviation for a comparison between the two Sr clocks. After an integration time of one hour, the $\tau^{-1/2}$ trend reaches $7 \times 10^{-17}$, meaning that at least one of the clocks reaches a stability of $5 \times 10^{-17}$. Inset: single scan over the atomic resonance in Sr$_1$ using a Rabi interrogation of 250~ms duration. The Fourier limited width is $3.2\,$Hz, corresponding to a quality factor of $Q=1.3 \times 10^{14}$. The contrast is close to 90\,\%.}
	\label{fig:Rabi250ms}
\end{center}
\end{figure}

Here, we measure the vector and tensor term with percent uncertainty. We improve the upper limit on the hyperpolarizability effect by more than one order of magnitude, and finally perform the first experimental study of the M1/E2 effect showing that it is controllable to within $10^{-17}$ even for $U_0$ as large as $700\,E_r$. These measurements are performed by comparing two $^{87}$Sr clocks (Sr$_1$ and Sr$_2$ in the following). Both clocks share the same clock interrogation laser at 698\,nm~\cite{Millo09,ThesisPhilip}. Fig.~\ref{fig:Rabi250ms} shows a single scan over the resonance of Sr$_1$ with a Rabi interrogation of duration $T=250\,$ms. The resulting Fourier limited linewidth of $3.2\,$Hz corresponds to a quality factor of $Q=1.3\times 10^{14}$; one of the highest ever obtained~\cite{Boyd06,Rosenband07}. The Allan deviation of the clock comparison is displayed in Fig.~\ref{fig:Rabi250ms} and reaches a level of $7\times 10^{-17}$ after one hour of averaging time. This record mid-term stability together with the possibility to operate the clock at extremely high $U_0$, up to $U_0=900\,E_r$, are the key features of our apparatus for the study of lattice related effects.

The lattice light at 813\,nm is supplied by an all semi-conductor laser source. The setup consists of an extended cavity diode laser (ECDL), whose frequency is locked to a Fabry-Perot transfer cavity. The cavity itself is referenced to the \drain\ atomic transition at 689 nm. The locked ECDL frequency exhibits long term fluctuations in the MHz range limited by the fluctuations of the cavity dispersion. Over a few hours, however, which is the typical duration of a measurement, the frequency is stable to within a few hundred kHz. The light from the ECDL is sent to two separate semi-conductor based amplifying systems providing up to 1.3\,W to each of the lattice clocks. The high-power light is transmitted and spatially filtered through an optical fiber and coupled to a resonant cavity surrounding the atomic sample to create a 1D lattice trap with depths up to $900\,E_r$. Semi-conductor amplifiers are known to produce non-negligible incoherent light spanning over tens of nm around the carrier frequency, which is removed by an interference filter of spectral width $\sim 0.1\,$nm. We checked that the non-filtered incoherent light had a negligible effect on the clock frequency by changing the coherent/incoherent light intensity ratio for various lattice depths.

We first consider the shifts resulting from the electric dipole (E1) interaction. The clock frequency shift $\Delta \nu^{\textrm{E1}}$ can be split into scalar, vector and tensor contributions, which for a given $|F,m_F\rangle$ state can be written~\cite{PhysRevA.59.4547,ThesisPhilip}
\begin{equation}
\label{eq:vector_pol_full}
	\Delta \nu^{\textrm{E1}} = (\Delta \kappa^s +  \Delta \kappa^v m_F\,\xi\vec e_k \cdot \vec e_B + \Delta \kappa^t \beta)\,U_0,
\end{equation}
where $\beta = (3|\vec \varepsilon \cdot \vec e_B|^2-1) [3m_F^2 - F (F+1)]$, $\vec e_k$ and $\vec e_B$ are unitary vectors along the lattice wavevector and the quantization axis, respectivelty, $\vec \varepsilon$ is the complex polarization vector of the lattice light, and $\xi\vec e_k = i\vec \varepsilon\wedge \vec \varepsilon^{\,*}$ its degree of circularity.

For each data point presented below, three to four sequences with different lattice depths are interleaved to provide a determination of lattice related shifts. A fit of the motional sidebands provides the oscillation frequency in the trap as well as the longitudinal and transverse temperatures. From these, we determine the average trap depth $U_0$ over the thermal distribution of the atoms in the trap. For each value of $U_0$, measurements with the symmetrical $m_F=\pm 9/2$ states are alternated. Their degeneracy is lifted by applying a bias field $\vec B_0$.

\begin{figure}
\begin{center}
	\includegraphics[width=\columnwidth]{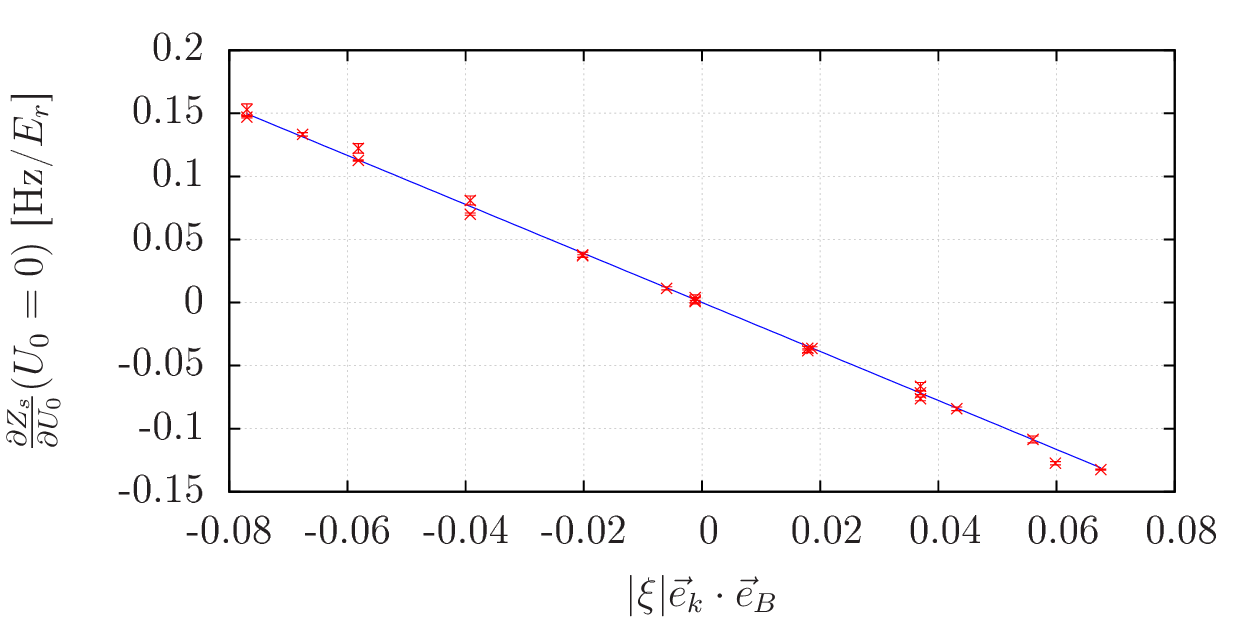}
	\caption{Linear component in $U_0$ of the Zeeman splitting between the $m_F = 9/2$ and $m_F = -9/2$ lines for different orientations of the quantization axis $\vec e_B$. This orientation is set by adding a small magnetic  field $\vec b$ parallel to the lattice, in addition to the usual orthogonal bias field $\vec B_0$ ($B_0 = 360\,\mu$T). The scalar product $\vec e_k \cdot \vec e_B \simeq \vec e_k \cdot \vec b / B_0$ is deduced from the Zeeman splitting $Z_s(U_0=0) \propto \sqrt{b^2 + B_0^2}$. As expected from (\ref{eq:vector_pol_full}), the dependence is linear and its slope gives $|9\Delta \kappa^v|$.}
	\label{fig:VectorGraph}
\end{center}
\end{figure}

The vector component in (\ref{eq:vector_pol_full}) is observed by measuring the frequency difference $Z_s = \nu(m_F) - \nu(-m_F)$ between two opposite $m_F$ states, which is insensitive to the scalar and tensor components. Its linear dependence on $U_0$ reflects the vector shift and its extrapolation to $U_0=0$ gives the first order Zeeman splitting. As expected from (\ref{eq:vector_pol_full}), the vector shift vanishes for a linear lattice polarization, as well as for a bias field orthogonal to the lattice. We perform an accurate determination of $\Delta \kappa^v$ by operating Sr$_2$ with a circular lattice polarization ($|\xi| = 1$) for various bias field orientations, using Sr$_1$ as a stable frequency reference. The results are displayed in Fig.~\ref{fig:VectorGraph}. A linear fit to these data yields~\footnote{The uncertainty is largely dominated by the systematic effects associated to the high atomic temperature due to the inefficiency of the narrow band cooling with a circular trap polarization.}
\begin{equation}
	|\Delta \kappa^v | = (0.22 \pm 0.05) \, \textrm{Hz}/E_r.
\end{equation}
Note that the vector shift does not directly affect the clock frequency, since it is rejected on average by alternately probing symmetrical Zeeman states. If the lattice polarization is linear, the vector shift also does not affect the bias field calibration, which must be done with percent accuracy to determine the frequency shift $\Delta\nu^{q}$ due to the second order Zeeman effect. We measured its the coefficient:
\begin{equation}
	\frac{\Delta\nu^{q}}{Z_s^2} = -0.246 (2)\,\textrm{Hz/kHz}^2, \textrm{or}\ \frac{\Delta\nu^{q}}{B_0^2} = -23.5 (2) \,\textrm{MHz/T}^2.
\end{equation}

\begin{figure}
\begin{center}
	\includegraphics[width=\columnwidth]{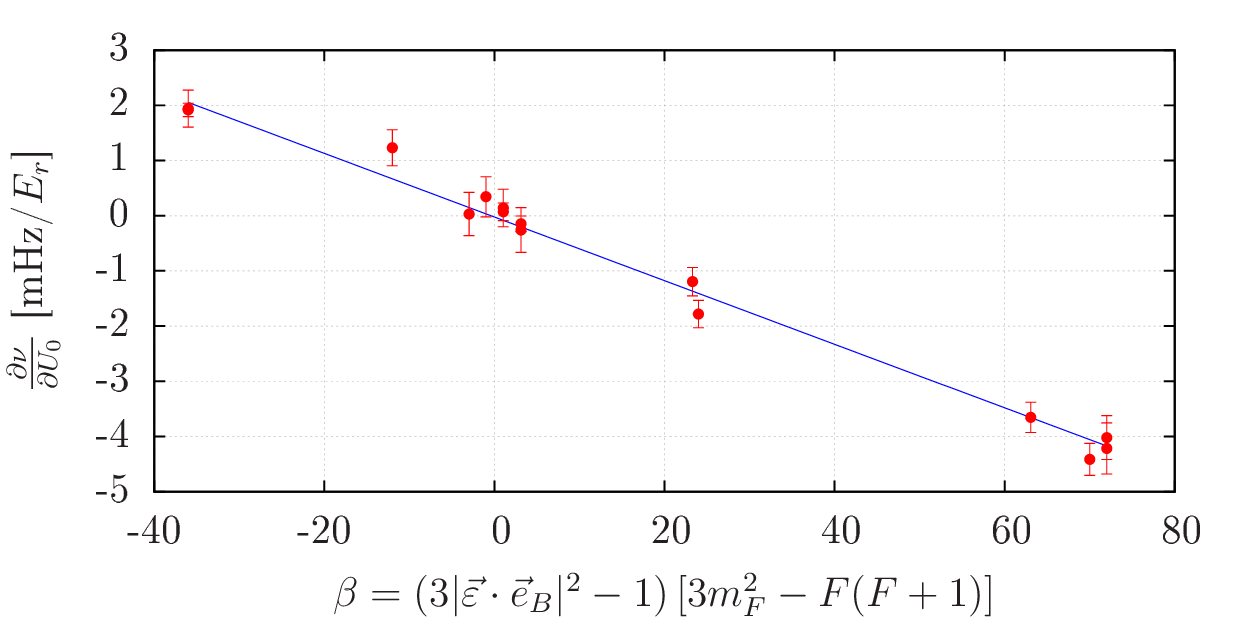}
	\caption{Linear dependence of the average clock frequency on the trap depth. We vary the $\beta$ parameter by changing the orientation of the quantization axis $\vec e_B$, by probing the $|m_F| = 7/2$ and $|m_F| = 9/2$ line and by switching between the two linear polarization eigen-modes of the lattice cavity. The lack of vector shift observed in Sr$_1$ ensures that the polarization of the lattice light is linear to a high degree. For $\beta$ = 0, we observe no dependence of the clock frequency on the trap depth, meaning that the lattice is tuned to the magic wavelength.}
	\label{fig:ExpTensorGraph}
\end{center}
\end{figure}

The linear component of the dependence of the average clock frequency $\nu \equiv \left(\nu(m_F) + \nu(-m_F)\right)/2$ on the trap depth $U_0$ is $(\Delta \kappa^s + \Delta \kappa^t \beta)U_0$, which gives access to the scalar and tensor polarizabilities. By varying the parameter $\beta$ on Sr$_1$ while using Sr$_2$ as a frequency reference (Fig.~\ref{fig:ExpTensorGraph}), we measure the tensor shift coefficient
\begin{equation}
	\label{eq:tensor_shift_coeffient}
	\Delta \kappa^t = (-0.0577\pm 0.0023) \, \textrm{mHz}/E_r.
\end{equation}
Its uncertainty is dominated by the knowledge of the atomic temperature, that determines the average power experienced by the atoms. This is the first observation of the tensor shift in a lattice clock. If the fluctuations of $\beta$ are kept smaller than 1, the induced clock frequency fluctuations due to the tensor shift will be below $10^{-17}$ for $U_0 = 100\,E_r$. This would correspond to fluctuations of $\vec \varepsilon \cdot \vec e_B$ at the $10^{-3}$ level which can easily be achieved for any lattice polarization/bias field configuration. By operating the experiment in a configuration with a second order polarization dependence of $\beta$ ($|\vec \varepsilon \cdot \vec e_B|^2=0$ or 1), $\beta$ can certainly be kept constant to within $0.1$ or less. This confirms that the tensor shift will not limit the accuracy of Sr lattice clocks.

Moreover, one can take advantage of the tensor shift to get rid of polarization inhomogeneities in a 3D lattices. This can be done by choosing different polarizations (hence different values of $\beta$) for the lattice beams propagating in different directions and accordingly tuning their frequencies so that the total linear shift $(\Delta \kappa^s + \Delta \kappa^t \beta)U_0$ is canceled everywhere in the lattice. The frequency difference would wash out any polarization interference between orthogonal directions, giving a perfectly homogeneous effective polarization in the lattice. If for instance one uses $|\vec \varepsilon \cdot \vec e_B|^2 = 0$ \emph{i.e.} $\beta = -36$ in two directions, and $|\varepsilon \cdot \vec e_B|^2 = 1$ \emph{i.e.} $\beta = 72$ in the third one, the frequency difference is 300 MHz and the configuration is such that the $\beta$ dependence on possible polarization fluctuations is second order for all the lattice beams.


Making $\beta = 0$ gives an accurate determination of the magic wavelength of the lattice for which the scalar term $\Delta \kappa^s$ in~\eqref{eq:vector_pol_full} is canceled. We find
\begin{equation}
	\nu_\textrm{m} = 368\, 554\, 693 (5) \, \textrm{MHz},
\end{equation}
which is in agreement with previous measurements \cite{Ludlow08}, with an accuracy improved by more than one order of magnitude.

\begin{figure}
\begin{center}
	\includegraphics[width=\columnwidth]{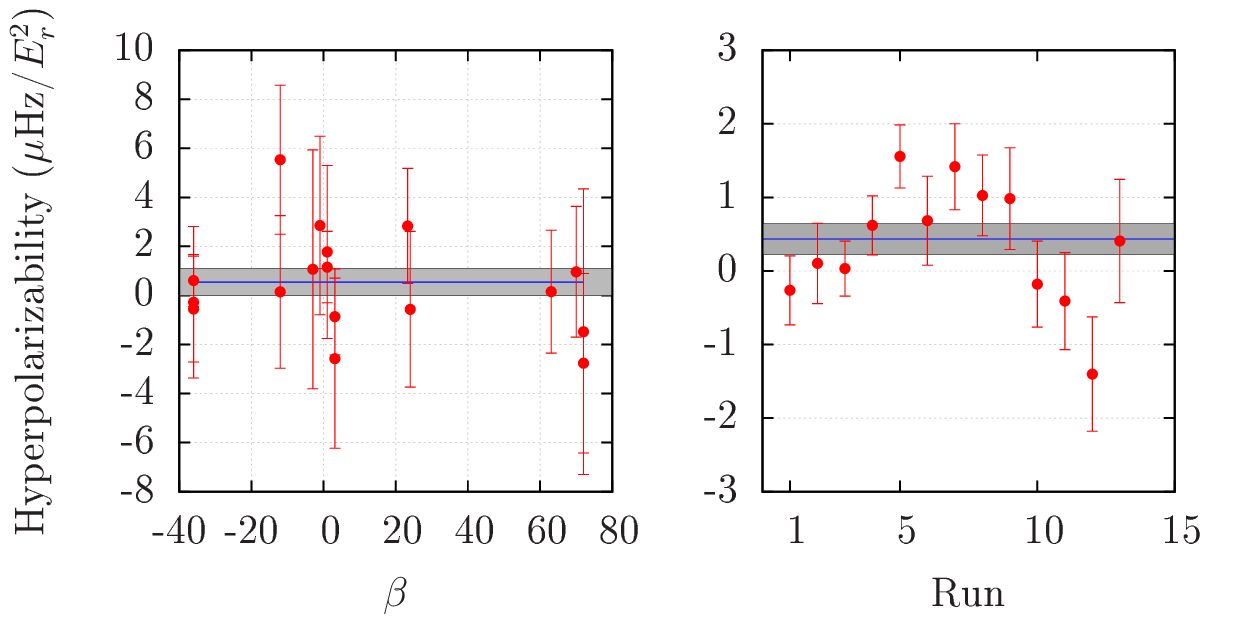}
	\caption{Left: hyperpolarizability extracted from the tensor shift data on Sr$_1$ shown in Fig.~\ref{fig:ExpTensorGraph}. Right: hyperpolarizability obtained from a preliminary measurement of the vector shift on Sr$_2$ with $|\xi| \sim 0.1$ for various orientations of $\vec e_B$. The latter exhibits a higher resolution due to a deeper trap. The grayed areas show the 1-$\sigma$ deviation from the weighted mean.}
	\label{fig:Hyperpol_exp}
\end{center}
\end{figure}

The hyperpolarizability is the $U_0^2$ dependence of the average clock frequency $\nu$. Its coefficient was extracted by fitting the same data that gave the vector and tensor shift measurements with a parabola, and is plotted in Fig.~\ref{fig:Hyperpol_exp}. Though it is not resolved here, the hyperpolarizability shift is expected to depend on the lattice geometry~\cite{taichenachev_optical_2006}. However, whatever this dependence, there exists at least one configuration where the effect is smaller than the weighted mean of the two datasets:
\begin{equation}
	\label{eq:hyperpol_weighted_mean}
	\Delta \nu^\textrm{Hyper.} = (0.46 \pm 0.18 )\, \mu \textrm{Hz} \, (U_0/E_r)^2.
\end{equation}
Its uncertainty is a factor of 60 lower than the previous evaluation reported in~\cite{Brusch06}, and is at the level of $10^{-17}$ or lower if the lattice depth is kept at $U_0 \leq 150 E_r$.

\begin{figure}
\begin{center}
	\includegraphics[width=\columnwidth]{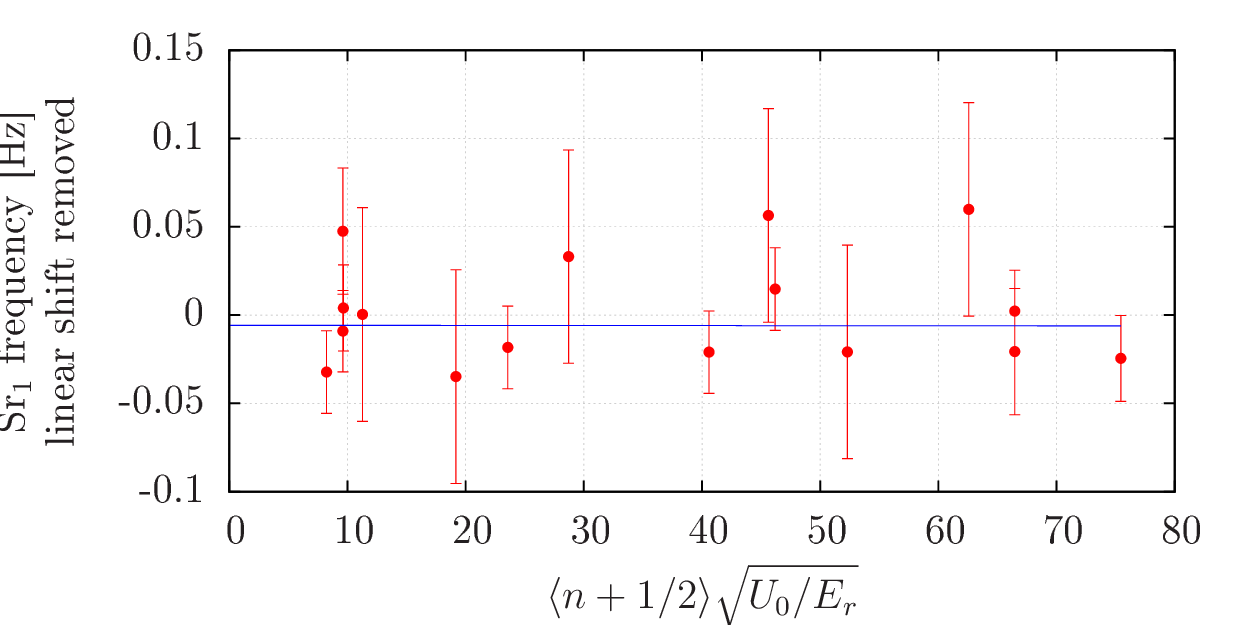}
	\caption{M1/E2 dependence of the average clock frequency $\nu$ of Sr$_1$. We observe no dependence of the clock frequency, but can put an upper bound on the $\zeta$ coefficient.}
	\label{fig:M1E2_shift}
\end{center}
\end{figure}

For evaluating the frequency shift due to the M1/E2 term, we use both the dependence on $U_0$ and on the atomic motional state in the lattice. Since M1 and E2 interactions modify the oscillation frequency in each lattice well, the shift of the clock transition for an atom in motional state $|n\rangle$ can be written
\begin{equation}
	\label{eq:M1E2_Shift}
	\Delta \nu^\textrm{M1/E2} = \zeta (n+1/2)\sqrt{U_0}.
\end{equation}
Eq.~\ref{eq:M1E2_Shift} holds for a harmonic trapping potential, which is a good approximation for the large lattice depths and low atomic temperatures considered here. Experimentally $n$ is varied by changing the atomic temperature $T$ along the strong confinement direction while keeping the transverse temperature constant. The effective quantum number is then $\mean{n}=\left( \exp(h\nu_l/k_B T)-1 \right)^{-1}$, where $\nu_l$ is the oscilation frequency along the lattice axis. Measurements are performed by interleaving configurations with different lattice depth and temperatures, with $U_0$ and $\mean{n}$ ranging from 50 to 400\,$E_r$ and from 0.5 to 3.5 respectively. The data are displayed in Fig.~\ref{fig:M1E2_shift}, giving the coefficient
\begin{equation}
	\zeta = (0 \pm 0.31)\,\textrm{mHz}/\sqrt{E_r}.
\end{equation}

The shift is not resolved here but gives a very stringent upper limit on the effect in optimal operating conditions.
For the coolest configuration achived here ($\mean{n} = 0.5$) and a trap depth of $150\,E_r$, the  uncertainty of the M1/E2 shift is smaller than $10^{-17}$. We therefore conclude that this effect is much smaller than first feared and does not constitute a threat to the ultimate performance of Sr clocks.

In conclusion, we have evaluated all relevant shifts due to the trap for a Sr lattice clock, showing that they will not thwart the ultimate performance of the clock. Sr clocks can even be operated at relatively large potential depths ($U_0 \sim 150\,E_r$ or more) while still keeping the lattice related perturbations below $10^{-17}$. We also propose using the tensor shift for designing a 3D lattice configuration with no polarization gradients and a perfect cancelation of the linear shift.

We thank Y. Le Coq for the magic wavelength measurements. SYRTE is a member of IFRAF (Institut Francilien de Recherche sur les Atomes Froids). This work has received funding from the European Community's FP7, ERA-NET Plus, under Grant Agreement No. 217257, as well as from CNES and ESA (SOC project).


\end{document}